\documentclass[12pt]{article}
\usepackage{latexsym}
\begin{document}
\title{ \vspace{0.5cm} Renormalizable elektroweak model without fundamental scalar
mesons.}
\author{A.A.Slavnov.\\Steklov Mathematical Institute
\\Gubkina st. 8, 119991 Moscow, \\Moscow State University}
\maketitle

\begin{abstract}
A renormalizable model of electroweak interaction which coincides
with Weinberg-Salam model in the gauge boson - fermion sector but
does not require the existence of fundamental scalar fields is
proposed.
\end{abstract}

\section{Introduction}

The Weinberg-Salam model \cite{We}, \cite{Sa} gives an excellent
description of electroweak processes in the framework of
renormalizable field theory. The essential ingredient of this
model is the scalar field providing spontaneous breaking of $SU(2)
\times U(1)$ symmetry via Higgs mechanism \cite{Hi}, \cite{Ki},
which allows to generate a mass term for the vector boson without
breaking the gauge invariance.

Predictions of the Weinberg-Salam model for the fermion-vector
meson sector are in a very good agreement with experiment. However
all attempts to find experimentally the scalar Higgs meson up to
now failed. Of course it is possible that such meson will be found
on LHC or other big machines. Nevertheless it seems worth to look
for an alternative model which preserves good predictions of the
Weinberg-Salam model in the fermion-gauge meson sector but does
not require existence of an elusive spin zero boson. Such a model
is proposed in this paper.

\section{Higher derivative reformulation of the Higgs-Kibble model.}

We start with reformulation of the usual $SU(2)$ Higgs-Kibble
model which is suitable for generalization described further. Such
a formulation was given in our paper \cite{Sl1}. It is based on
the higher derivative action
\begin{eqnarray}
A= \int dx \{ L_{YM}+(D_{\mu}\varphi)^+(D_{\mu}\varphi)+
\nonumber\\
\frac{g}{2m}\partial_{\mu}X\partial_{\mu}(\varphi^+\varphi)+
\frac{a^{-2}}{2}\Box X \Box X+\partial_{\mu} \bar{c}
\partial_{\mu}c \}
 \label{1}
\end{eqnarray}
Here $L_{YM}$ is the $SU(2)$ Yang-Mills Lagrangian, $\varphi$ is
the complex doublet with the components
\begin{equation}
\varphi_1= \frac{iB_1+B_2}{\surd{2}}; \quad \varphi_2= \frac{
\surd{2}m}{g}+ \frac{1}{\surd{2}}(\sigma-iB_3) \label{2}
\end{equation}
$m$ is the mass which the Yang-Mills quanta acquire via Higgs
mechanism, $D_{\mu}$ is the usual covariant derivative and
$\bar{c}, c$ are anticommuting ghost fields.

Under the gauge transformations the Yang-Mills field $A_{\mu}$ and
the scalar fields $\varphi$ transform in a standard way and the
fields $X$ and $ \bar{c}, c$ are singlets of the gauge group. Due
to the presence of the constant term in the component $\varphi_2$,
the quadratic part of this action is not diagonal with respect to
the fields $X$ and $\sigma$. To diagonalize it we make a shift
$\sigma \rightarrow \sigma-X$. The action (\ref{1}) acquires a
form
\begin{eqnarray}
A=\int dx \{\tilde{L}(A_{\mu},B,\tilde{\sigma})+ \frac{1}{2}
\partial_{\mu}\sigma \partial_{\mu} \sigma- \frac{1}{2}
\partial_{\mu}X\partial_{\mu}X+ \nonumber\\
\frac{a^{-2}}{2} \Box X \Box X+ \partial_{\mu} \bar{c}
\partial_{\mu}c+ \frac{g}{4m} \partial_{\mu}X \partial_{\mu}
(B^2+ \tilde{\sigma}^2) \} \label{3}
\end{eqnarray}
Here $ \tilde{L}+ \frac{1}{2} \partial_{\mu} \sigma \partial_{\mu}
\sigma$ is the usual Lagrangian for the Yang-Mills field
interacting with $B$ and $\sigma$, in which $\sigma$ in the
interaction is replaced by $ \tilde{\sigma}= \sigma-X$.

It was shown in our paper \cite{Sl1} that quantization of this
model with the help of Ostrogradsky canonical formalism in the
gauge $B=0$ leads to the following spectrum: three massive spin
one particles associated with the fields $A_{\mu}$, and one spin
zero massless field $\sigma$. The field $X$ generates two states:
one massive spin zero particle with the mass $a$ and one massless
spin zero state with negative norm. The corresponding propagators
look as follows
\begin{equation}
D^{ab}_{\mu\nu}= \frac{g^{\mu\nu}-k^{\mu}k^{\nu}m^{-2}}{k^2-m^2}
\delta^{ab} \label{4}
\end{equation}
\begin{equation}
D_{\sigma}= \frac{1}{k^2} \label{5}
\end{equation}
\begin{equation}
D_X= \frac{a^2}{k^2(k^2-a^2)}\label{6}
\end{equation}

Due to invariance of the action (\ref{3}) with respect to the
supersymmetry transformations
\begin{eqnarray}
\delta X=c \varepsilon; \quad \delta \sigma=c \varepsilon
\nonumber\\
\delta \bar{c}=X \varepsilon- \sigma \varepsilon-a^{-2} \Box X-
\frac{g}{4m}(B^2+ \tilde{\sigma}^2) \label{7}
\end{eqnarray}
there exists a conserved operator $Q$, which separates the
invariant subspace
\begin{equation}
Q|\Phi>=0 \label{8}
\end{equation}
with the nonnegative norm \cite{Sl1}. The explicit solution of eq.
(\ref{8}) has a form
\begin{equation}
|\Phi>=[1+\sum_n c_n \prod_{1\leq j \leq
n}(\sigma^+(k_j)-a^+(k_j))]|\Phi>_{A,b} \label{9}
\end{equation}
where $ \sigma^+$ and $a^+$ are the creation operators of the
positive and negative norm massless states respectively, and $
|\Phi>_{A,b}$ describes the states with the massive vector field
and the scalar particle with the mass $a$.

The spectrum of observables in this model is identical to the
Higgs model, and it is easy to show that the transition amplitudes
also coincide. Explicitly renormalizable perturbation theory may
be constructed by passing from the gauge $B=0$ to some
renormalizable gauge like the Lorentz gauge
$\partial_{\mu}A_{\mu}=0$.

It is worth to note that in distinction of the standard Higgs
model, where in the gauge $B=0$ only physical states are present
and physical unitarity is manifest, in our formulation even in the
$B=0$ gauge unphysical states are present and the unitarity is
achieved by imposing on the physical vectors the condition
(\ref{8}).

The unitarity of the model is easily understood if one looks at
the structure of the $X$-field propagators (\ref{6}):
\begin{equation}
\tilde{D}_X= \frac{a^2}{k^2(k^2-a^2)}= \frac{1}{k^2-a^2}-
\frac{1}{k^2}
\label{10}
\end{equation}
The field $X$ enters the interaction in the action (\ref{3})
either in combination $ \tilde{\sigma}= \sigma-X$ or in the form
$\Box X$. In the first case the contribution of the negative norm
component of $X$ exactly compensates the contribution of the
massless excitation $\sigma$, and in the second case due to the
presence of D'Alambert operator the negative norm excitations are
not created in the asymptotic states.

Although the physical content of the model described above is
equivalent to the Higgs model, it is more convenient for
generalizations including additional scalar particles. Contrary to
the Higgs model where the physical scalar particles belong to
multiplets transforming nontrivially under the gauge
transformations, in our formulation the physical scalar particle
is the gauge singlet. It allows to introduce additional terms
depending on $X$-fields without breaking the gauge invariance of
the model.

\section{A new family of renormalizable vector meson models
with spontaneously broken symmetry.}

Let us consider the model described by the following action
\begin{eqnarray}
A=\int dx[L_{YM}+(D_{\mu} \varphi)^+(D_{\mu} \varphi)+
\frac{g}{2m}
\partial_{\mu}(\sum_{i=1-N/2}^{N/2}X_i)\partial_{\mu}(\varphi^+
\varphi)+\nonumber\\
\frac{1}{2}(\sum_{i=1-N/2}^{N/2} \partial_{\mu}X_i)^2- \frac{N}{2}
\sum_{i=1-N/2}^{N/2}
\partial_{\mu}X_i \partial_{\mu}X_i+ \frac{N}{2}
\sum_{i=1-N/2}^{N/2}a_i^{-2} \Box X_i \Box X_i] \label{11}
\end{eqnarray}
We did not write here the ghost fields $\tilde{c}, c$ which are
free and do not influence the results. The complex doublet field
$\varphi$ is parameterized as in eq.(\ref{2}).

Having in mind that the fields $X_i$ are singlets of the gauge
group we see that the action (\ref{11}) is gauge invariant.

To diagonalize the quadratic part of this action we again make a
shift
\begin{equation}
\sigma \rightarrow \sigma-\sum_{i=1-N/2}^{N/2}X_i \label{12}
\end{equation}
Then the action (\ref{11}) acquires a form
\begin{eqnarray}
A= \int dx \{\tilde{L}(A_{\mu},B, \tilde{\sigma})+ \frac{1}{2}
\partial_{\mu}
 \sigma \partial_{\mu} \sigma-
 \frac{N}{2} \sum_i \partial_{\mu}X_i \partial_{\mu}X_i +\nonumber\\
\frac{N}{2} \sum_i a_i^{-2} \Box{X_i} \Box{X_i}+  \frac{g}{4m}
\partial_{\mu}( \sum_i X_i) \partial_{\mu}(B^2+ \tilde{\sigma}^2)
\label{13}
\end{eqnarray}
Here as in the eq.(\ref{3}) $\tilde{L}+ \frac{1}{2} \partial_{\mu}
\sigma \partial_{\mu} \sigma$ is the usual Lagrangian for the
Yang-Mills field interacting with $B$ and $\sigma$ with the
substitution $\sigma \rightarrow \tilde{\sigma}$.
\begin{equation}
\tilde{\sigma}= \sigma-\sum_{i=1-N/2}^{N/2}X_i
 \label{14}
\end{equation}
Imposing the gauge condition $B^a=0$ one easily finds the
propagators of the fields $A_{\mu}, \sigma, X_i$. The propagators
of $A_{\mu}$ and $ \sigma$ are given by the eqs.(\ref{4},
\ref{5}), and the propagators of $X_i$ are equal to
\begin{equation}
\tilde{D}_{ij}^X(k)= \delta_{ij} \frac{a_i^2}{Nk^2(k^2-a_i^2)}
\label{15}
\end{equation}
Hence the propagators of $\tilde{\sigma}$ are
\begin{equation}
\tilde{D}_{\tilde{\sigma}}(k)= \frac{1}{k^2}+ \sum_{i=1-N/2}^{N/2}
\frac{a_i^2}{Nk^2(k^2-a_i^2)}= \sum_{i=1-N/2}^{N}
\frac{1}{N(k^2-a_i^2)} \label{16}
\end{equation}
One sees that the propagators of $\tilde{\sigma}$ have no zero
mass poles. The field $\sigma$ exactly compensates the
contribution of zero mass components of $X$ and only positive norm
components with the masses $a_i$ are propagating. As in the
previous model the interaction terms include the fields $X_i$ only
in the combinations $\sigma-\sum_i X_i$ or $ \Box X_i$. Neither of
these combinations generates negative norm states and the theory
is unitary in the physical subspace including only massive vector
fields and the massive components of the fields $X_i$.

The interaction of the scalar particles is suppressed by the
factor $(N)^{-1}$ and for large $N$ the probability of creation of
a zero spin particle with the fixed mass $a_i$ is very small. At
the same time in the intermediate states all massive components of
$X_i$ fields contribute and for $|k|\gg a_i$, the propagator
$\tilde{D}_{\tilde{ \sigma}}(k)$ coincides with the usual Higgs
meson propagator. However for $k \sim a_i$ the interaction of
scalar particles may differ considerably from the Higgs meson
interaction.

Obviously one can choose instead of the gauge $B^a=0$ some other
gauge condition, for example the Lorentz gauge
$\partial_{\mu}A_{\mu}=0$. In this gauge the vector field
propagator decreases at $k \rightarrow \infty$ as $k^{-2}$ and
there are additional $B$-field propagators, which also decrease as
$k^{-2}$. In this gauge the model is manifestly renormalizable.

As we discussed before for large values of $N$ creation of one
particle spin zero states is suppressed, but contribution of all
scalar particles produces a collective effect which replaces the
Higgs meson exchange in the Weinberg-Salam model. This collective
effect may be attributed to the manifestation of a special extra
dimension.

Indeed, the model described above may be considered as having a
discrete extra dimension. The Yang-Mills fields and the fields
$\varphi$ are living on the four-dimensional "brane", whereas the
fields $X_i$ and their masses depend on the fifth coordinate.

Continuous version of such a model may be described by the action
\begin{eqnarray}
A=\int dx[L_{YM}+(D_{\mu}\varphi)^+(D_{\mu}\varphi)+\nonumber\\
\frac{g}{2m}\partial_{\mu}(\varphi^+ \varphi) \int_{-
\pi/\kappa}^{\pi/\kappa} \partial_{\mu}X(x, \lambda) d \lambda+
\frac{1}{2} \int_{- \pi/\kappa}^{\pi/\kappa}
\partial_{\mu}X(x, \lambda) d \lambda \int_{-\pi/\kappa}^{\pi/\kappa}
\partial_{\mu}X(x, \lambda) d \lambda+\nonumber\\
\frac{\pi}{\kappa} \int_{-\pi/\kappa}^{\pi/\kappa} \Box X(x,
\lambda) \Box X(x, \lambda)a^{-2}(\lambda) d \lambda-
\frac{\pi}{\kappa} \int_{-\pi/\kappa}^{\pi/\kappa}
\partial_{\mu}X(x, \lambda) \partial_{\mu}X(x, \lambda) d\lambda]
\label{18}
\end{eqnarray}
where $\varphi$ is again parameterized as in eq.(\ref{2}).

Diagonalizing the quadratic part of the action (\ref{18}) by the
shift
\begin{equation}
\sigma(x) \rightarrow \sigma(x)-
\int_{-\pi/\kappa}^{\pi/\kappa}X(x, \lambda) d \lambda \label{19}
\end{equation}
we get
\begin{eqnarray}
A= \int dx[L(A_{\mu},B, \tilde{\sigma})+ \frac{1}{2}
\partial_{\mu} \sigma \partial_{\mu} \sigma+\nonumber\\
\frac{g}{4m} \partial_{\mu}(B^2+ \tilde{\sigma}^2)
\int_{-\pi/\kappa}^{\pi/\kappa} \partial_{\mu}X d \lambda-
\frac{\pi}{\kappa} \int_{-\pi/\kappa}^{\pi/\kappa}
\partial_{\mu}X \partial_{\mu}X d \lambda+\nonumber\\
\frac{\pi}{\kappa} \int_{-\pi/\kappa}^{\pi/\kappa}\Box X \Box
Xa^{-2}(\lambda) d \lambda] \label{20}
\end{eqnarray}
Here
 \begin{equation}
\tilde{\sigma}(x)= \sigma(x)- \int_{-\pi/\kappa}^{\pi/\kappa}X(x,
\lambda)d \lambda \label{21}
\end{equation}

The $X$-field propagator defined by the action (\ref{20}) is
\begin{equation}
\tilde{D}(k, \lambda, \mu)= \delta(\lambda-\mu)
\frac{\kappa}{2\pi} \frac{a^2(\lambda)}{k^2(k^2-a^2(\lambda))}
\label{22}
\end{equation}
Discretized version corresponds to $ \frac{2\pi}{\kappa}=Nb$, where
$b$ is the lattice spacing.

The propagator of the field $ \tilde{\sigma}$ looks as follows
\begin{eqnarray}
\tilde{D}_{\tilde{\sigma}}(k)= \frac{1}{k^2}+
\int_{-\pi/\kappa}^{\pi/\kappa} d \lambda d \mu
\frac{\kappa}{2\pi}[-\frac{1}{k^2}+ \frac{1}{k^2-a^2(\lambda)}]
\delta(\lambda-\mu)=\nonumber\\
\frac{\kappa}{2\pi} \int_{-\pi/\kappa}^{\pi/\kappa} d \lambda
\frac{1}{k^2-a^2(\lambda)} \label{24}
\end{eqnarray}
It has no pole at $k^2=0$ and for $k^2 \gg a^2$
$\tilde{D}_{\tilde{\sigma}}$ coincides with the Higgs meson
propagator.

As an example one can take
\begin{equation}
a^2(\lambda)= \lambda+ \frac{\pi}{\kappa}+M_0^2, \quad \lambda<0;
\quad a^2(\lambda)=- \lambda+ \frac{\pi}{\kappa}+M_0^2, \quad
\lambda>0 \label{25}
\end{equation}

Then
\begin{equation}
\tilde{D}_{\tilde{\sigma}}(k)=- \frac{\kappa}{\pi} \ln(1-
\frac{\pi}{\kappa (k^2-M_0^2)}) \label{26}
\end{equation}

The propagators of the fields $\tilde{\sigma}$ and $ \Box X$ have
no one particle pole, but generate a branch point corresponding to
the collective excitations. One sees that for $k^2 \gg
\frac{\pi}{\kappa}+M_0^2$ the function
$\tilde{D}_{\tilde{\sigma}}$ coincides with the usual Higgs meson
propagator, $ \tilde{D} \sim k^{-2}$, but for small $k^2$ the
behavior of $ \tilde{D}_{\tilde{\sigma}}$ is quite different.

\section{Electroweak model.}

The mechanism described in the previous section may be applied
directly to the description of electroweak interactions. The
fermion-gauge meson sector coincides with the Salam-Weinberg
model, but the Higgs sector is described by the Lagrangian
\begin{eqnarray}
L= |\partial_{\mu}\varphi+ig \frac{\tau^a}{2}A_{\mu}^a \varphi-
\frac{ig_1}{2}B_{\mu} \varphi|^2-G( \bar{L} \varphi R+ \bar{R}
\varphi L+ \ldots)
\nonumber\\
+ \frac{g}{2m} \sum_i \partial_{\mu}X_i
\partial_{\mu}(\varphi^+ \varphi)+ \frac{1}{2} \sum_i
\partial_{\mu}X_i \sum_i \partial_{\mu}X_i-\nonumber\\
\frac{N}{2} \sum_i \partial_{\mu}X_i \partial_{\mu}X_i+
\frac{N}{2} \sum_i a_i^{-2} \Box X_i \Box X_i \label{27}
\end{eqnarray}
where summation over $i$ goes from $1-N/2$ to $N/2$. Letters $L,R$
denote lepton doublets and singlets and $ \ldots$ stand for the
corresponding terms with quark fields. The field $\varphi$ is
parameterized as before
\begin{equation}
\varphi_1= \frac{(B_1+iB_2)}{\surd{2}}; \quad
\varphi_2=\frac{\surd{2}m}{g}+ \frac{\sigma-iB_3}{\sqrt{2}}
\label{28}
\end{equation}
and $A_{\mu}^a, B_{\mu}$ are the gauge fields corresponding to
$SU(2)$ and $U(1)$ subgroups.

The Lagrangian (\ref{27}) is gauge invariant if the fields
$A_{\mu}, B_{\mu}, \varphi, L, R$ transform in a usual way and the
$X_i$ fields are the gauge group singlets.

The constant component of $\varphi_2$ generates mass terms for
gauge fields and fermions and produces mixing between the fields
$X_i$ and $\sigma$. Making the shift (\ref{14}) we obtain the
model whose gauge boson - fermion sector is identical to the
Weinberg-Salam model, and the scalar mesons are described by the
action (\ref{11}) with obvious modifications due to the presence
of the $U(1)$ gauge field $B_{\mu}$. The interaction with fermions
enters via the terms
\begin{equation}
G[ \bar{L}( \varphi- \sum_iX_i)R+h.c.+ \ldots]  \label{29}
\end{equation}
In the limit $N \rightarrow \infty$ the model may be described by
the action of the type (\ref{18}) with continuous extra dimension.
In this case one particle poles describing spin zero particles
completely disappear from the spectrum and instead there is a
branch point corresponding to some collective excitations. For a
finite $N$ our model describes the system with several neutral
spin zero particles.

\section{Conclusion.}
In this paper we described the renormalizable gauge invariant
model of massive Yang-Mills field, which does not require the
existence of fundamental scalar particles. A discretized version
describes a gauge invariant model of massive vector field with
spontaneously broken symmetry and several neutral scalar mesons.
This mechanism may be applied to the electroweak model of the
Salam-Weinberg type to modify the predictions concerning spin zero
particles. We considered here in details only the $SU(2)$ group as
being the most important for electroweak models, but all the
results are trivially generalized to other gauge groups.

{\bf Acknowledgements} \\
This work was partially supported by RFBR under grant 050100541,
grant for support of leading scientific schools 20052.2003.1, and
the program "Theoretical problems of mathematics".
 \end{document}